\documentstyle[prl,aps]{revtex}
\begin{document}
\twocolumn[
\title{Scaling behaviour in daily air humidity fluctuations}
\author{G\'abor Vattay\cite{LAbs}}
\address{Niels Bohr Institute,
Blegdamsvej 17, DK-2100  Copenhagen \O, Denmark}
\author{Andrea Harnos}
\address{University of Veterinary Science, Department of Biomathematics
and Computer Science, Istv\'an u. 2, H-1078 Budapest, Hungary}
\date{\today}

\maketitle

\mediumtext
\begin{abstract}
We show that the daily average air humidity fluctuations exhibit
non-trivial $1/f^{\alpha}$ behaviour which different from
the spectral properties of other meteorological quantities.
This feature and the fractal spatial structure found in
clouds make it plausible to regard air humidity fluctuations as a
manifestation of self-or\-ga\-ni\-zed cri\-ti\-ca\-li\-ty.
We give arguments why the dynamics in air humidity
can be similar to those in sandpile models of SOC.
\end{abstract}
\pacs{05.40.+j,47.25.-c, 44.25.+f,92.60.-e}
\narrowtext
]
There is widespread interest in understanding the dynamic processes of large,
spatially extended systems. There are many geophysical processes in
nature where scale-invariance is observed. Recently, the
Gutenberg-Richter law of earthquakes\cite{eq}, the volcanic
activity\cite{va} and various astrophysical phenomena\cite{astro}
have been reported to show scale-invariant properties.
However, such quantitative analysis is often limited by a shortage of
available data or a lack of understanding of the basic effects.
One of the largest extended systems is our atmosphere.
Although all the basic mechanisms that govern the
dynamics of the atmosphere have been well known for quite a long time,
a detailed understanding and adequate characterization of the fluctuations
of various statistical quantities of the lower atmospheric boundary
layer are still not complete. The conventional viewpoint is that
the underlying mechanism is random and can be considered as
an autoregressive process. Quite surprisingly, in Ref.\cite{JanVat}
it was possible to demonstrate that the distribution
and spectral properties of the daily medium temperature fluctuations
coincide with those measured in a helium cell in Refs.\cite{KadLib,HesLib}.
The daily mean temperature fluctuations
on a one day scale and the {\em Soft Turbulent} 2K helium cell
fluctuations on a millisecond scale look the same since
the dimensionless parameters, the Prandtl and Raylegh numbers
of the two systems, are the same.
The main objection to using daily average data for deeper
analysis is the belief that random weather changes (fronts, etc.)
dominate the temporal behaviour of the quantities measured at
a typical meteorological station. In reality, in the case of daily
temperature fluctuations, it was observed that the meteorological
fronts due to
the large scale coherent spatial motion affect only the short (1-10
days)
timescale, while the vast majority of timescales (10-10000 days)
is governed by the inherent fluctuations of the Benard-Raylegh
convection in the lower 500m boundary layer\cite{JanVat}.
In this Letter we would like to show that the fluctuations of the
daily average relative air humidity data also reflect some
inherent physical processes and that this measurement is an ideal candidate
for observing large scale-invariance in the atmosphere.

The air humidity can be quantified by the partial vapor pressure $p_v$
of gaseous water in the air.
At a given temperature $T$, the amount of water in the air is limited.
The maximum partial vapor pressure corresponding to the maximum
water content is the saturation vapor pressure $p_{sat}(T)$.
The relative air humidity $e$ at temperature $T$ is the ratio of the
actual to the saturation pressure measured in percent
\begin{equation}
e=\frac{p_v}{p_{sat}(T)}\cdot 100\% .
\end{equation}
A detailed summary of the measurement and the definition of the
air humidity and the saturation vapor pressure
are given in the classic work of Penman\cite{Penman}.

The special importance of the humidity from the point of view
of self-or\-ga\-ni\-zed cri\-ti\-ca\-li\-ty (SOC)\cite{BTW}
lies in its threshold behaviour.
If the actual humidity in the air locally exceeds $100\%$, the
water starts condensing. In the absence of macroscopic spatial
motion of the air, the microscopic amount of condensed
water is transported on microscopic scales by diffusion.
It can reach a new location, where the local relative humidity
is under $100\%$ and can return to the gaseous state.
If the humidity in the neighbourhood is also saturated,
the amount of condensed water will increase. This can initiate a
chain reaction like the avalanches in the `sandpile' models\cite{BTW}.
We propose that large decreases in the air humidity are
consequences of such processes.
On the other hand, the humidity in the air is coming
from open water surfaces, from the moisture in soil and
from plants and animals. If the air humidity is under $100\%$,
the evaporation from these sources continually increases the water
content of the air. This is the analog of dropping sand
on the sandpile, and this drives the system toward the fully saturated
$100\%$ humidity state which is the stationary and critical state of the
system at the same time. It is very likely that these aspects of the
dynamics help to build up a self-or\-ga\-ni\-zed cri\-ti\-ca\-l state on
large scales. The two main characteristics of the SOC behaviour is the
$1/f^{\alpha}$ type behaviour of the power spectra of
the fluctuations, and the spatial scaling behaviour manifested in the
formation of fractal structures\cite{BTW}.
One of the most well known fractals in nature are
clouds\cite{Mand}, which are formed in the chain reaction-like
condensation of air humidity. These processes and large scale fractal
behaviour of clouds can be observed in meteorological satellite
pictures\cite{sate}. The condensation of air
humidity into clouds can be modeled by cellular automata, which
has recently been reported to show SOC behaviour\cite{Nagel}.

To detect scaling in the temporal behaviour, we have analyzed relative
air humidity data collected by the Hungarian Meteorological Service
in various stations in Hungary during the period
1963 - 1988. The main difficulty in analyzing meteorological
time series is that seasonal periodic variations can cause
peaks in the Fourier transforms and make it difficult
to determine the power spectra of the fluctuations.
Figure 1 shows the seasonal variation
$\bar{e}(i)$ of a typical station (Balaton, county Heves),
which is defined as
\begin{equation}
\bar{e}(i)=\frac{1}{26}\sum_{y=1963}^{1988}e(i,y), \;\; i=1,...,365
\label{svar}
\end{equation}
where $e(i,y)$ denotes the daily average air humidity of the $i$th
day of the $y$th year measured at the meteorological station.
To diminish the effect of seasonal variations we subtract it
from the daily data.
Then the fluctuation $e_f$ is given by the deviation of the actual
humidity from the seasonal variation (\ref{svar}):
\begin{equation}
e_f(i,y)=e(i,y)-\bar{e}(i),
\label{fluct}
\end{equation}
where the notation is the same as above.
In this way we obtain a $26\times 365$ day long record of
daily average relative air humidity fluctuations.
Hungary can be considered as meteorologically homogeneous from
the point of view of fluctuations\cite{Har} due to the
geography. Different stations measure coherent
fluctuations\cite{JanVat},
and one station can be considered as representative.
Figure 2 shows a typical record of the daily average humidity
fluctuations for the year 1969.
Figure 3 shows a histogram of the daily average humidity
fluctuations, which is clearly a Gaussian distribution,
with zero mean and standard deviation of $\sigma=9.34\%$.
The power spectrum $S(f)$,which is the squared magnitude of
the Fourier amplitude, can be obtained by standard numerical
methods\cite{Numrec}. Figure 4 shows the power spectrum
of the time series measured in Balaton. Almost the whole frequency
range can be fitted by a simple power law
\begin{equation}
S(f)\sim \frac{1}{f^{\alpha}},
\end{equation}
with $\alpha=0.61\pm 0.01$.
In addition to the large scale spatial scaling behaviour this shows
the existence of scaling behaviour also in the time domain. A very nice
feature of this result is that meteorological fronts and
other uncorrelated events do not seriously affect the spectra.
If the underlying process is SOC, the exponent $\alpha$ can be
related to the scaling of the weighted average of the
avalanches\cite{Chr}. The weighted average is defined as
\begin{equation}
\Lambda(t)=\sum_s s^2 P(s,t),
\end{equation}
where $P(s,t)$ is the distribution of the avalanche sizes $s$
as a function of the duration $t$ of the avalanches.
For SOC this scales as
\begin{equation}
\Lambda(t)\sim t^{\mu},
\end{equation}where the exponent satisfies\cite{Chr} $-1<\mu<+1$. The
exponents
$\mu$ and $\alpha$ are related\cite{Chr} as $\alpha=\mu+1$,
which yields $\mu=-0.39\pm0.01$.

Finally, to demonstrate the non-trivial nature of a
$1/f^{\alpha}$ type power-density spectrum,
on Figures 5 and 6 we show the power spectra of
the daily temperature and rain fluctuations respectively, measured at the
same meteorological station. The temperature record
has been analyzed in Ref.\cite{JanVat}, while the rain record
appears to be white noise (ie. a constant power
spectrum). Neither show any sign of scaling behaviour.

In summary, we have found that the spectral density of
daily average air humidity fluctuations shows scaling
behaviour. This is very different from
the spectral properties of other meteorological quantities.
This and the fractal behaviour of the condensed air humidity in
space make it very plausible to regard air humidity fluctuations as a
manifestation of self-or\-ga\-ni\-zed cri\-ti\-ca\-li\-ty.
We gave arguments why the dynamics in air humidity
can be similar to those in sandpile models of SOC.
We hope that this study can inspire more detailed analysis
of the space-time dynamics of clouds and air humidity fluctuations.

We wish to thank Zs. Harnos for inspiration
and the records of daily average humidity, I. M. J\'anosi, T. Bohr,
M. H. Jensen and
K. Sneppen for useful comments and P. Dimon for careful reading of the
manuscript. This work was supported by OTKA under Grant No. F4286.
One of the authors (G. V.) thanks
the Sz\'echeny Foundation and the Danish Natural Science Research
Council for finantial
support and the hospitality of the CATS in the Niels Bohr Institute.

\begin{figure}
\caption{The seasonal variation of the daily average air humidity
(Eq.(2)) obtained from the data of the meteorological station
Balaton, county Heves, Hungary for the period 1963-1988.}
\end{figure}
\begin{figure}
\caption{Air humidity fluctuations (Eq.(3)) measured in 1969.}
\end{figure}
\begin{figure}
\caption{Unnormalized histogram of the air humidity fluctuations
measured in the period 1963-1988.
The dashed line is the fitted Gaussian distribution with standard deviation
$\sigma=9.34\%$}
\end{figure}
\begin{figure}
\caption{Unnormalized power spectrum of the daily average
relative air humidity
fluctuations. The dashed
line is the best fit given by (Eq.(4)).}
\end{figure}
\begin{figure}
\caption{Unnormalized power spectrum of the daily medium temperature
fluctuations.}
\end{figure}
\begin{figure}
\caption{Unnormalized power-density spectrum of the daily rain
fluctuations.}
\end{figure}

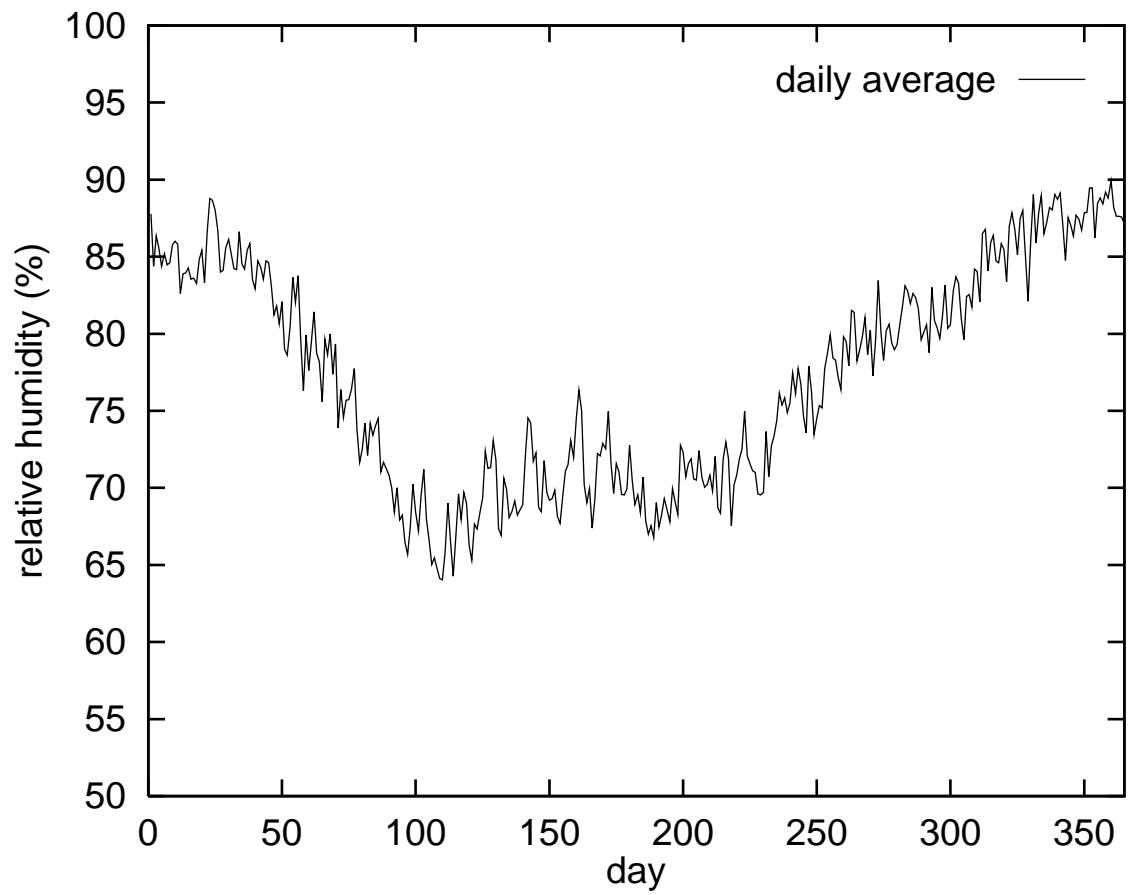
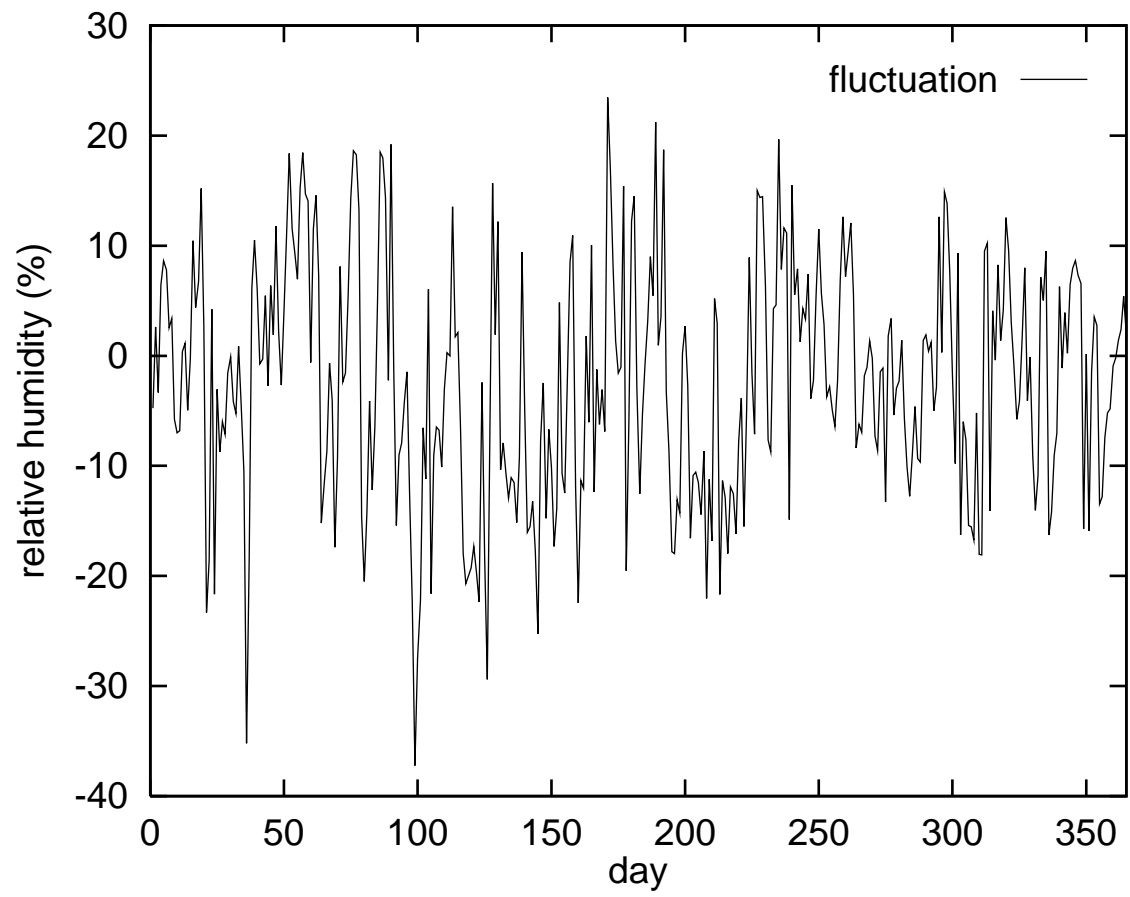
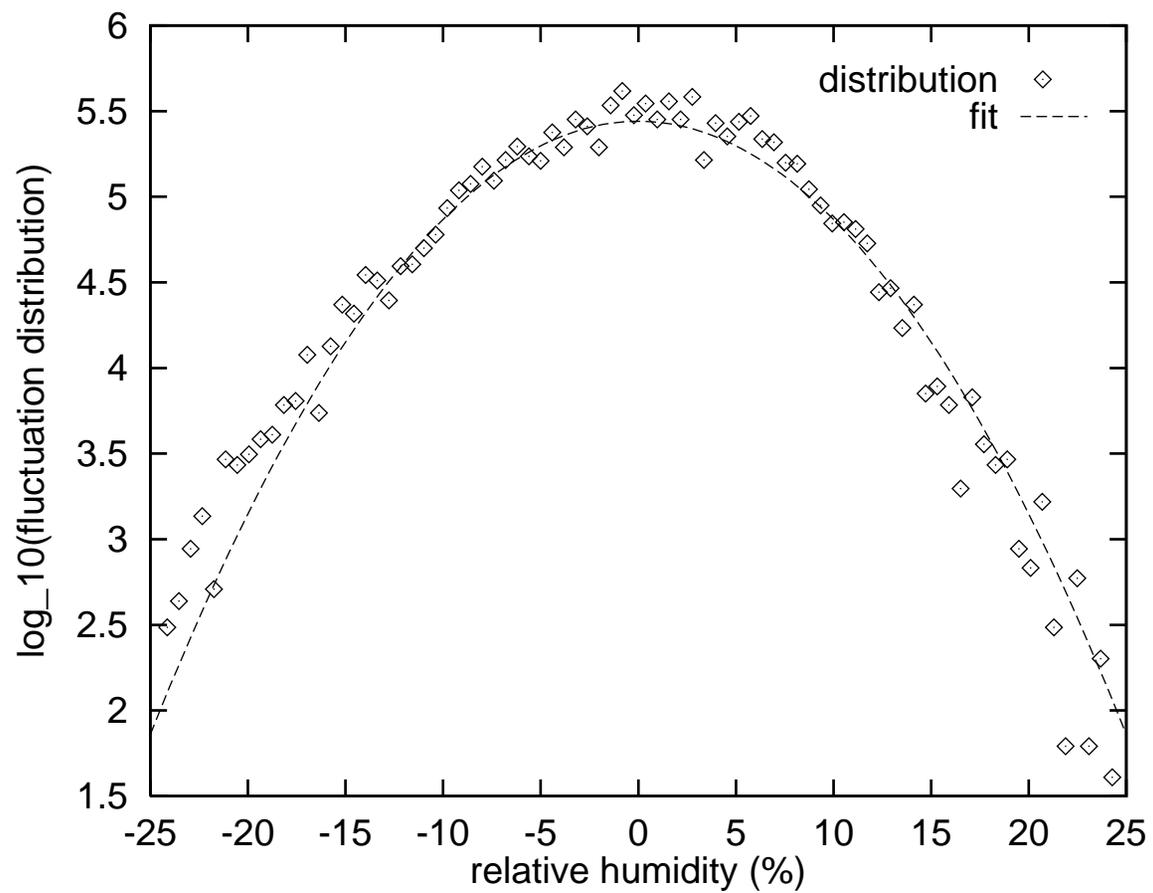
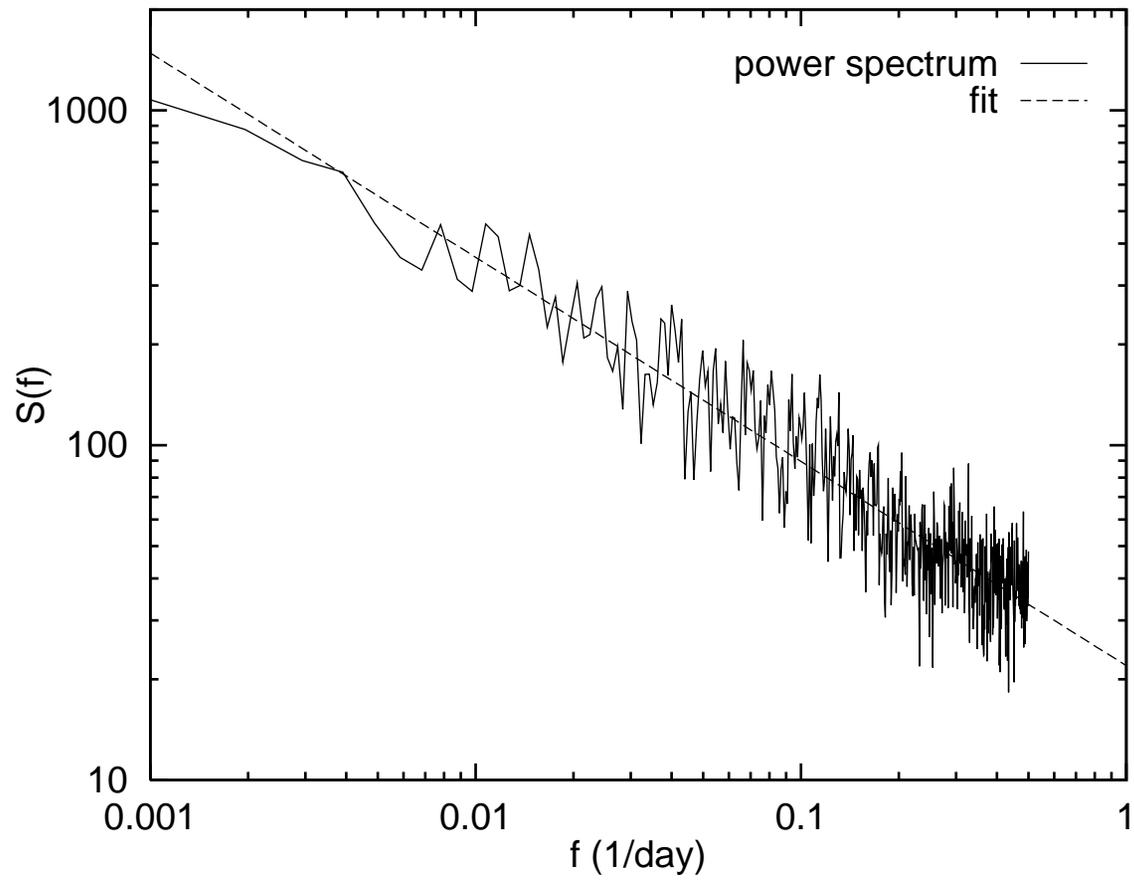
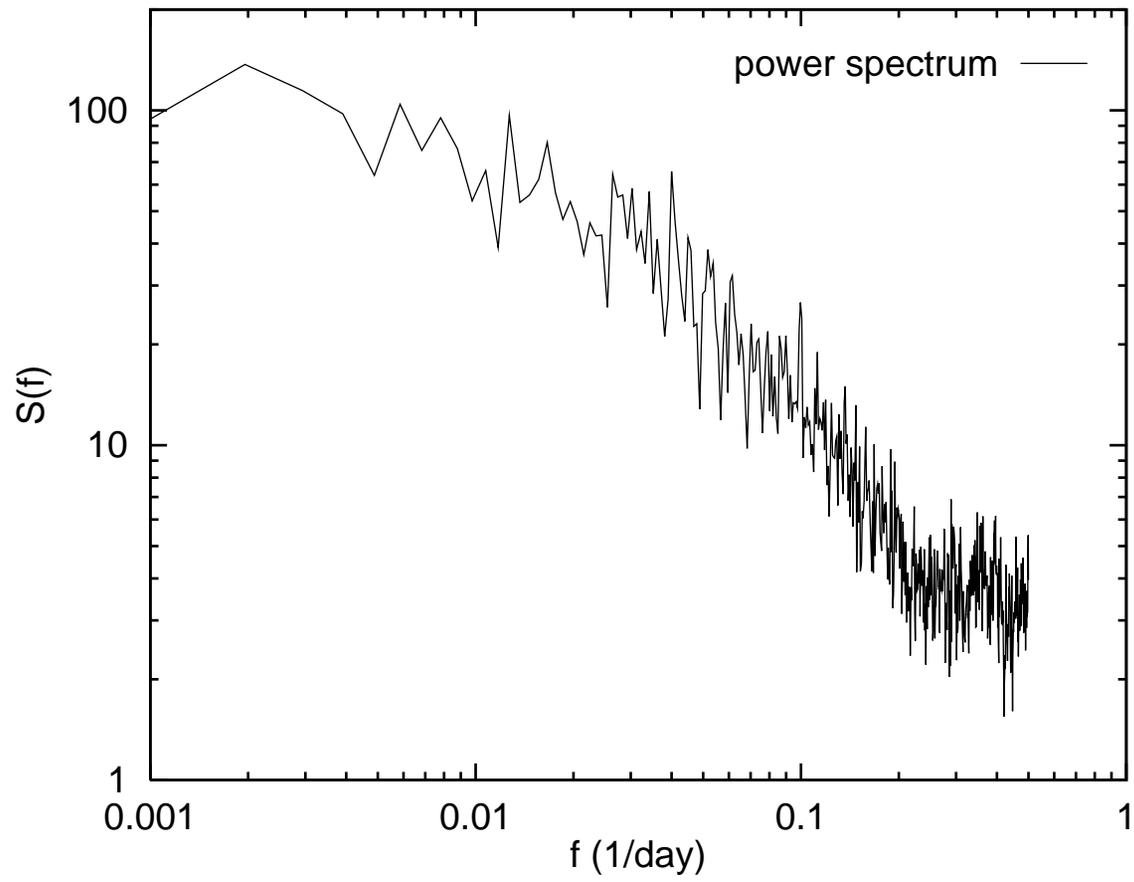
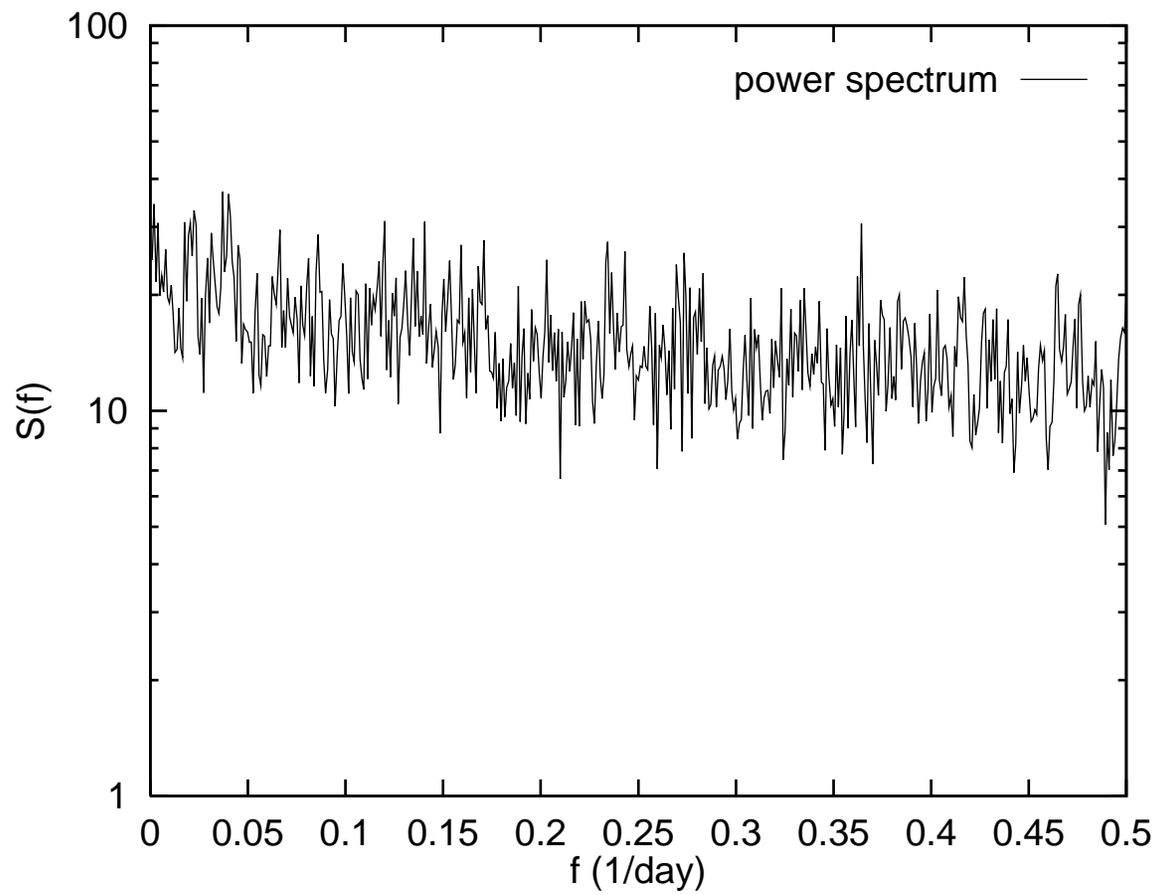
\end{document}